# A force torsor analysis for a turning process in the presence of self-excited vibrations

Constantin ISPAS[a], Olivier CAHUC[b], Claudiu BISU[a], Alain GERARD[b]

a University Politehnica Bucarest, 313 Splaiul Independentei, 060042 Bucharest, Roumanie
b Université de Bordeaux, 351, cours de la Libération, 33405 Talence cedex, France

**Abstract**
A testing device in turning including, in particular, a six-component dynamometer, is used to measure the complete torque of the cutting actions, in a case of self-excited vibrations. For the tests, the used tool was a noncoated carbide tool (TNMA 160412) without chip breaker. The cutting material is a chrome molybdenum alloy type (ASI 4140). The cylindrical test tubes have a diameter of 120 mm and a length of 30 mm. For the first time, we present an analysis of forces and moments for different depths of cut and different feed rates.

# 1 Introduction

In the three-dimensional cutting case, the torsor of mechanical actions (forces and moments) is often truncated: the moments part of this torsor is neglected for lack of adapted metrology [1, 2, 3]. Unfortunately, until now the results on the cutting forces are almost still validated using platforms of forces (dynamometers) measuring the three components of those. The torsor of actions is thus often truncated because the torsor moment part is probably neglected for lack of access to an adapted metrology [4, 5]. However, forces and pure moments (or torque) can be measured [6]. Recently, an application consisting in six-component measurements of the actions torsor during cutting process was carried out for the case of high speed milling [7], drilling [8, 9], etc. Cahuc et al., in [10], present another use of this six-component dynamometer in an experimental study: taking into account of the cut moments allows a better machine tool power consumption evaluation. It allows a better approach of the cut [8, 11, 12] and should thus allow in the dynamic case to reach new properties of the vibrations of the system piece-tool-matter.

Moreover, the tool torsor has the advantage of being transportable in any point of space and in especially, at the tool tip in O point. The following study is carried out in several stages, including two major stages; the first of is related to the analysis of forces. The second of is dedicated to the determining of the central axis and a first moments analysis to the central axis during the cut.

In paragraph 2 we present first the experimental device used and the associated elements of measurement. Paragraph 3 is devoted to the measurement of the torque of the cutting actions. An analysis of the forces exerted during the cut action is carried out. It allows to establish in experiments certain properties of the resultant of the cutting actions. The case of the moments at the tool tip point is also examined with precision. The central axis of the torque is required (paragraph 4). The beams of central axes deduced from the multiple tests confirm especially the presence of moments at the tool tip point. In paragraph 5, we more particularly carry out the analysis of the moments at the central axis by looking at the case the most sensitive to vibrations (ap = 5 mm, f = 0.1 mm/rev). Before concluding, this study gives a certain number of properties and drive to some innovative reflexions.

## 2  Experimental device

The experimental device presented on Figure 1 is a conventional lathe (Ernault HN 400). The machining system dynamic behaviour is identified using one three-direction accelerometer fixed on the tool. Two unidirectional accelerometers are positioned on the lathe, on the front bearing of the spindle, to identify the influence of this one during the cutting process. All the cutting actions (forces and torques at the tool tip point) are measured by a six-component dynamometer [6]. The instantaneous spindle speed is continuously controlled (with an accuracy of 1%) by a rotary encoder directly coupled with the workpiece. The connection is carried out by a rigid steel wire, which allows a better behaviour. The test workpieces are cylindrical, with a diameter of 120 mm and a length of 30 mm. The dimensions of these test tubes were selected using a finite element analysis coupled to an optimisation, with SAMCEF® software in order to confer on the unit a maximum rigidity. This procedure is described in [13]. Thus, under a load P = 1,000 N, for material having a Young modulus $E = 21.10^5$ N/mm², the workpiece dimensions selected are: $D_1 = 60$ mm (diameter), $L_1 = 180$ mm (length) for a bending stiffness of $7.10^7$ N/m (Figure 2). These values are within the higher limit of the rigidity zone acceptable for a conventional lathe [14, 15, 16].

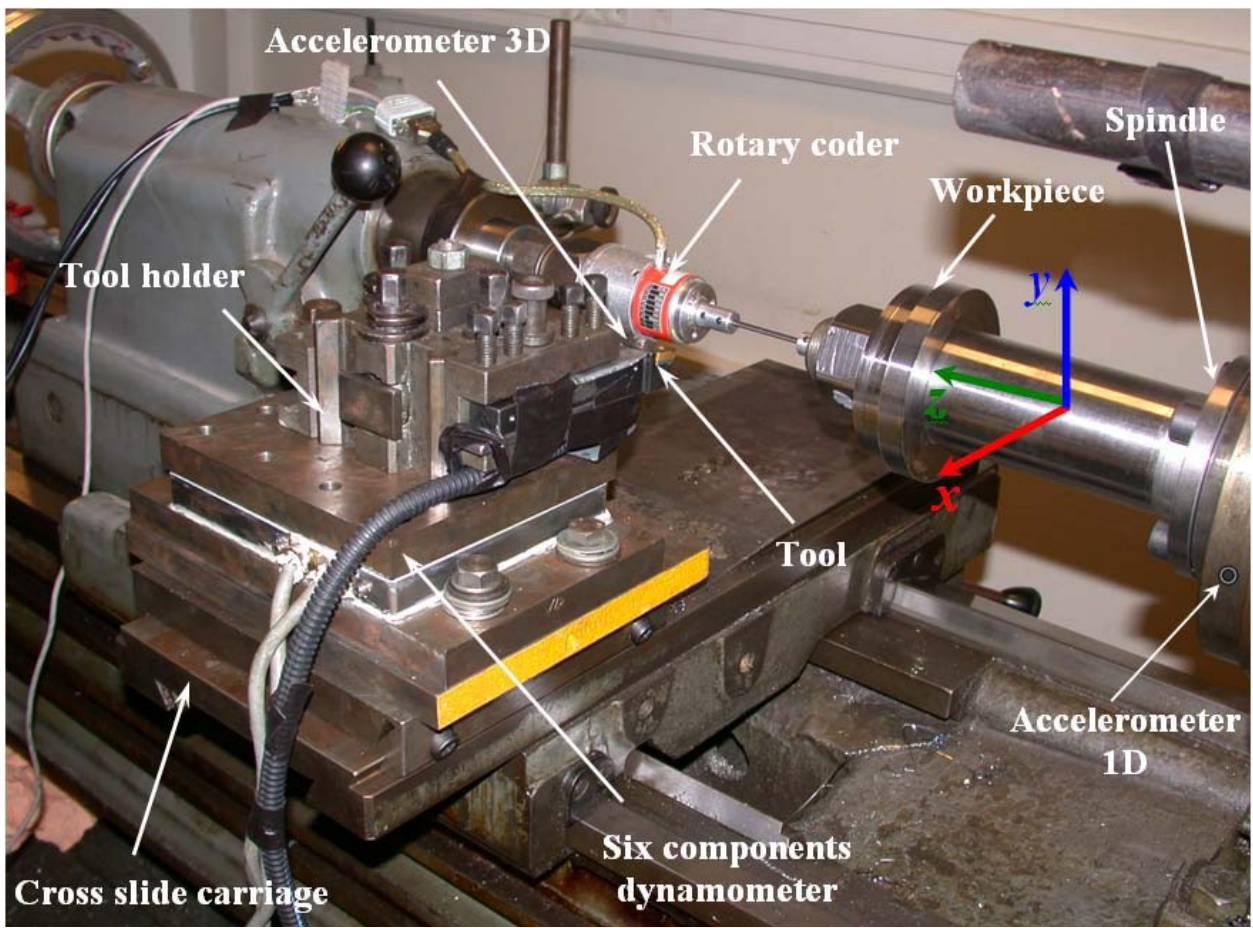

Figure [1]: Experimental device and associated measurement elements.

During the tests, the type of the insert tool used is TNMA 16 04 12 carbide not covered, without chip breaker. The machined material is an alloy of chrome molybdenum type 42CrMo24. Moreover, the tool geometry is characterized by the cutting angle $\gamma$, the clearance angle $\alpha$, the inclination angle of edge $\lambda_s$, the direct angle $\kappa_r$, the nozzle radius $r_\varepsilon$ and the sharpness radius R [17]. In order to limit to the wear appearance maximum along the cutting face, the tool insert is examined after each test, and is changed if necessary (Vb ≤ 0.2 mm ISO 3685). The tool parameters are detailed in the Table 1.

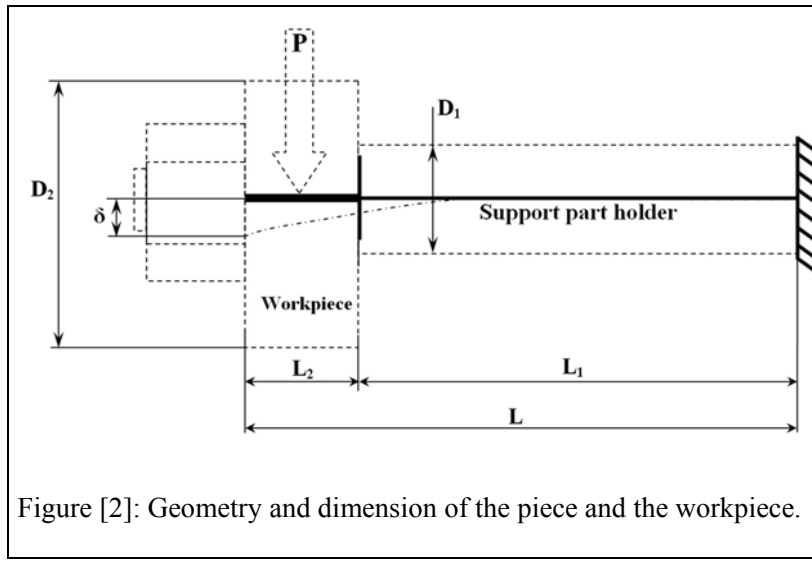

Figure [2]: Geometry and dimension of the piece and the workpiece.

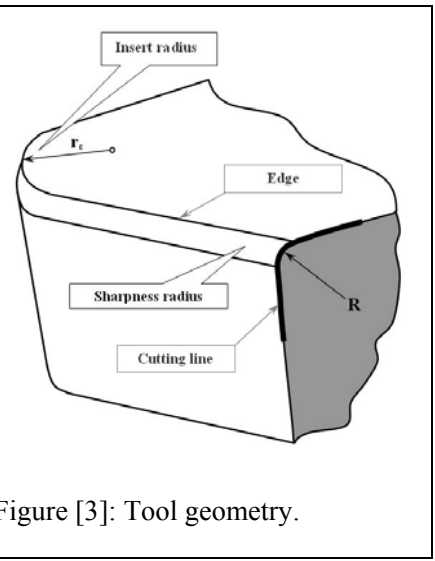

Figure [3]: Tool geometry.

| $\gamma$ | $\alpha$ | $\lambda$ | $\kappa_r$ | $r_\varepsilon$ | R |
|---|---|---|---|---|---|
| -6° | 6° | -6° | 91° | 1.2 mm | 0.02 mm |

Tableau [1]: Geometrical characteristics of the tool.

## 3 Cutting torsor actions

### 3.1 Tests

The experiments are performed within a framework similar to the one describe in Cahuc et al., [10]. For each test, the complete torsor of the mechanical actions is measured using the six-component dynamometer according to the method initiated in Toulouse [18], developed by Couétard [11], and used on several occasions [7, 10, 19, 20, 21]. These mechanical actions are evaluated for four depths of cut ap (= 1 mm; 2 mm ; 3.5 mm ; 5 mm) and in according to four feed rates f (= 0.1; 0.05; 0.0625; 0.075 mm/rev). The six-component dynamometer give the instantaneous values of all the torque cutting components in the three-dimensional space (x, y, z) related to the machine tool (Figure 1). Measurements are made in the six-component dynamometer transducer O' center and then transported to the tool point O via the moment transport traditional relations. Uncertainties of measurement of the six-component dynamometer are: ±4% for the forces components and ±6% for the moment components.

## 3.2 Resultant of the cutting actions analysis

For the four values feed rate f indicated above, two examples of resultant efforts measurements applied at the tool tip point are presented: one of these for the stable case (quasi without vibrations) ap = 2 mm and another for the case with instability (self-excited vibrations) ap = 5 mm. The Figure 4 shows signals related to the resultant components of cutting forces following the three (x, y, z) cutting directions. For the test case presented here, parameters used are: ap = 2 mm, f = 0.1 mm/rev and N = 690 rpm. The Figure 5 shows signals related to the resultant components of cutting forces following the three (x, y, z) cutting directions. For the test case here presented, parameters used are: ap = 5mm, f = 0.1 mm/rev and N = 690 rpm.

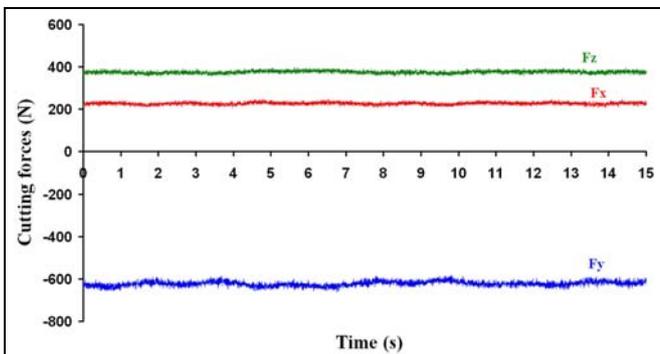
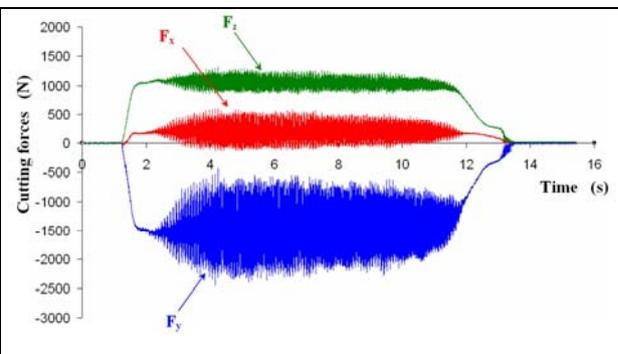

Figure [4]: Without vibrations signals related to the resultant components of cutting forces.

Figure [5]: With vibrations signals related to the resultant components of cutting forces.

In the stable case (ap = 2 mm, f = 0.1 mm/rev), it appears (Figure 4) that the force component amplitudes remain almost independent of the time parameter. Thus, the amplitude variation is limited to 1 or 2 N around their nominal values, starting with 200 N for ($F_x$), 400 N for ($F_z$), and 600 N for ($F_y$). These variations are quite negligible and lower than uncertainties of measurements. Indeed, when the nominal stress is reached, the component noticed as the lowest value is ($F_x$), while the highest in absolute value is ($F_y$). While taking as reference the absolute value of ($F_x$), the following relation between these three components comes:

$$|Fx| = |Fz|/ 2 = |Fy|/ 3 \text{ (stable case, ap=2 mm).} \quad (1)$$

In the unstable case (Figure 5, ap = 5 mm, f = 0.1 mm/rev), we observe that the force component on the cutting axis ($F_y$) has the most important average amplitude (1,500 N). It is also the most disturbed (±1,000 N) with oscillations between −2,450 N and −450 N. In the same way, the force along the feed rate axis ($F_z$) has important average amplitude (1,000 N), and the oscillations have less width in absolute value (±200 N), but also in relative value (±20%). As for the effort on the radial direction ($F_x$), it is weakest on average (200 N), but also most disturbed in relative value (±240 N). These important oscillations are the tangible consequence of the contact tool/workpiece frequent ruptures and, thus, demonstrate the vibration and dynamical behaviour of the system **WTM**.

Finally, taking as reference the absolute value of ($F_x$), the following relation between the absolute values of these three components comes:

$$|Fx| = |Fz|/ 5 = |Fy|/ 7.5 \text{ (unstable case, ap = 5 mm).} \quad (2)$$

In other words, whatever the depth of cut ap we have the following order relation between the absolute values of the cutting actions components:

$$|Fx| \leq |Fz| \leq |Fy|. \qquad (3)$$

More precisely, in the case of turning here studied, whatever the depth of cut ap = 2 mm (stable case) or ap = 5 mm (unstable case), we have, the following relation between the absolute values of the cutting actions components:

$$|Fx| = |Fz|/ap = |Fy|/1.5ap. \qquad (4)$$

with the errors of experiments close. So, the cutting actions are found proportional to the depth of cut, as in [22, 23].

In the unstable case (ap = 5mm, f = 0.1 mm/rev) the component of the force resultant highlights a plan in which a variable cutting forces Fv moves around a nominal value Fn [23]. This variable force is an oscillating action (Figure 6) that generates tool tip displacements and maintains the vibrations of elastic system block-tool **BT** [23]. Thus, the cutting force variable and the self-excited vibrations of elastic system **WTM** are interactive, in agreement with research work [5, 15, 24, 25].

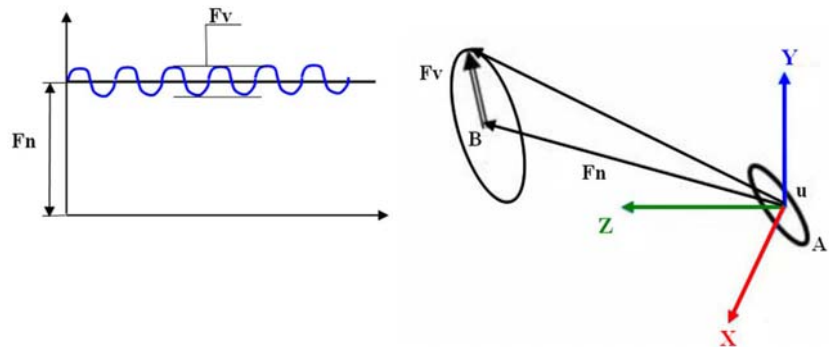

Figure [6]: Cutting force $F_v$ evolution around the nominal value $F_n$.

### 3.3 Study of the moments at the tool point

Just like the torsor resultant, we give two examples of statements of the moment components at the tool point in the machine frame (Figure 1). To compare, we present initially the stable case Figure 7 (without vibrations following the three directions of space machine in ap = 2 mm and f = 0.1 mm/rev), and then the unstable case Figure 8 (with vibrations following the three directions of space machine ap = 5 mm and f = 0.1 mm/rev.).

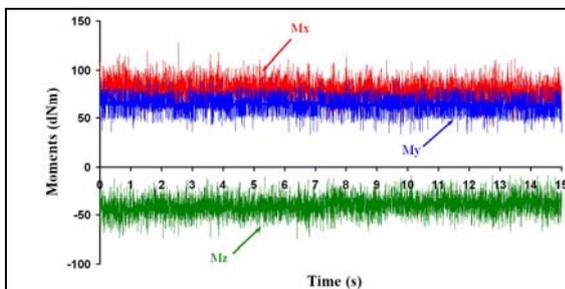 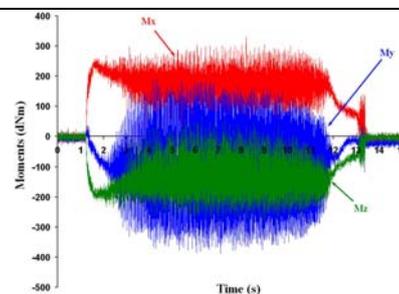

Figure [7]: The moment components time signals that act on the tool tip (ap = 2 mm).   Figure [8]: The moment components time signals that act on the tool tip (ap = 5 mm).

Just like the resultants in the stable case (Figure 7), the moment components average values at the tool tip point are slightly disturbed (quasi-constants except for some dN.m – déci Newton meter –) while in the unstable case (Figure 8) the average values of the moment components at the tool tip point are very disturbed. However, we can note that the moment components at the tool tip point are more disturbed than their equivalents regarding the resultant. Thus, the moment components seem more sensitive to the self-excited vibrations than the force components applied. The follow-up of those could thus be a good means of detecting the existence of regenerative vibrations precociously.

The analysis of the results shown on Figure 7 allows to establish that among the three moment components, the average value of the component along the axis in advance is lowest (Moz = – 40 dN.m). It is most disturbed in relative value (±20 dN.m i.e. 50%) while the absolute value of the component along the axis of cut plays the role of pivot (Moy = 60 dN.m) with a weaker disturbance in absolute value (±16 dN.m i.e. 30%). The highest component is along the radial axis (Mox = 80 dN.m) with a weaker disturbance in absolute value (±20 dN.m i.e. 25%). Taking as reference of the average values this of the component Moz, we have:

$$||Moz|| = ||Moy||/1.5 = ||Mox||/2. \qquad (5)$$

The analysis of the results (Figure 8) allows to establish that among the three moment components, the average value of the absolute value of the component along the axis of cut is the lowest (Moy = - 100 dN.m). It is also most disturbed (±280 dN.m i.e. variations of about 300%!). The component of the torque along the radial axis (x) is always positive Mox = 170 dN.m (±150 dN.m); it is thus the most raised moment component and, proportionally, the least disturbed (±75%). The Moz moment component according to the radial direction is always negative with an average value about - 140 dN.m in absolute value the highest but less disturbed than Moy (with oscillations of ±140 dN.m i.e. only ±100%). Finally, for this depth of cut, we have the following relation between the absolute values of the moment components:

$$||Moy|| = ||Moz||1,4 = ||Mox||/1,7 \quad (ap = 5 \text{ mm}), \qquad (6)$$

with a module of average couple at the tool tip point about $||Mo|| = 275$ dN.m. The comparison of Eq. (2) and Eq. (6), shows that the role of x and y axes is reversed for ap = 5 mm. This allows to think that the transport of the moments at the tool tip point has a major effect. Moreover, only the Mox modulus keeps the highest whatever the value of ap in the turning cases here considered. It is also the only positive and slightly decreasing component when the depth of cut increases. One notes that the components Moy and Moz are always negative when the depth of cut increases. The modulus of the moments component at the tool tip point, increases with the depth of cut, like the one of the resultant.

In the same way the general order relation (3) for the resultant is replaced for the moment with the tool tip point by:

$$|Moy| \leq |Moz| \leq |Mox|. \qquad (7)$$

The x and y axes position are also reversed there regarding to the resultant and moment at the tool tip point components. This can be allotted to the moments transport at the central axis. It is an additional incentive being studied of the moments at the central axis.

In addition, for the unstable case (ap = 5 mm, f = 0.1 mm/rev) the spatial evolution of the moments is no longer fully situated in a plane, as it is the case for forces. An 8, slightly out of plane is described (Figure 9) [27].

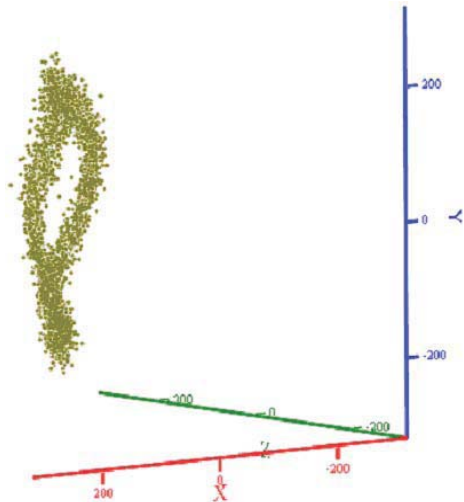
Figure [9]: Moments place space representation.

## 4  Central axis

It is well known that, with any torsor, it is possible to associate a central axis (except the torsor of pure moment), which is the single object calculated starting from the six torsor components [28]. A torsor $[A]_O$ in a point $O$ is composed of resultant forces **R** and the resulting moment $\mathbf{M_O}$

$$[A]_o = \left\{ \frac{\vec{R}}{\overrightarrow{M_o}} \right. . \tag{8}$$

The central axis is the line defined classically by:

$$\overrightarrow{OA} = \frac{\vec{R} \wedge \overrightarrow{M_o}}{\left|\vec{R}\right|^2} + \lambda \vec{R}, \tag{9}$$

where $O$ is the point where the mechanical action torsor was moved (here, the tool tip) and $A$ is the current point describing the central axis. Thus, **OA** is the vector associated with the bipoint $[O, A]$ (Figure 10). This line (Figure 10a) corresponds to geometric points where the mechanical actions moment torsor is minimal. The central axis calculation consists in determining the points assembly (a line) where the torsor can be expressed according to a slide block (straight line direction) and the pure moment (or torque) [28].

The central axis is also the point where the resultant cutting force is colinear with the minimum mechanical moment (pure torque). The test results enable us to check for each measurement point where the colinearity between the resultant cutting force **R** and moment **MA** calculated is related at the central axis (Figure 10b). The meticulous examination of the six mechanical action torsor components shows that the forces and the moment average values are not null. For each point of measurements, the central axis is calculated in the stable (Figure 11a) and unstable modes (Figure 11b). In any rigour, the case ap = 2 mm and f = 0.1 mm/rev should be described as quasi-stable movement because the vibrations exist but their amplitudes are very small —of the order of micrometers— thus, quasi null compared to the other studied cases. Considering the cutting depth value ap = 5 mm and f = 0.0625 mm/rev, the recorded amplitude was 10 times more important.

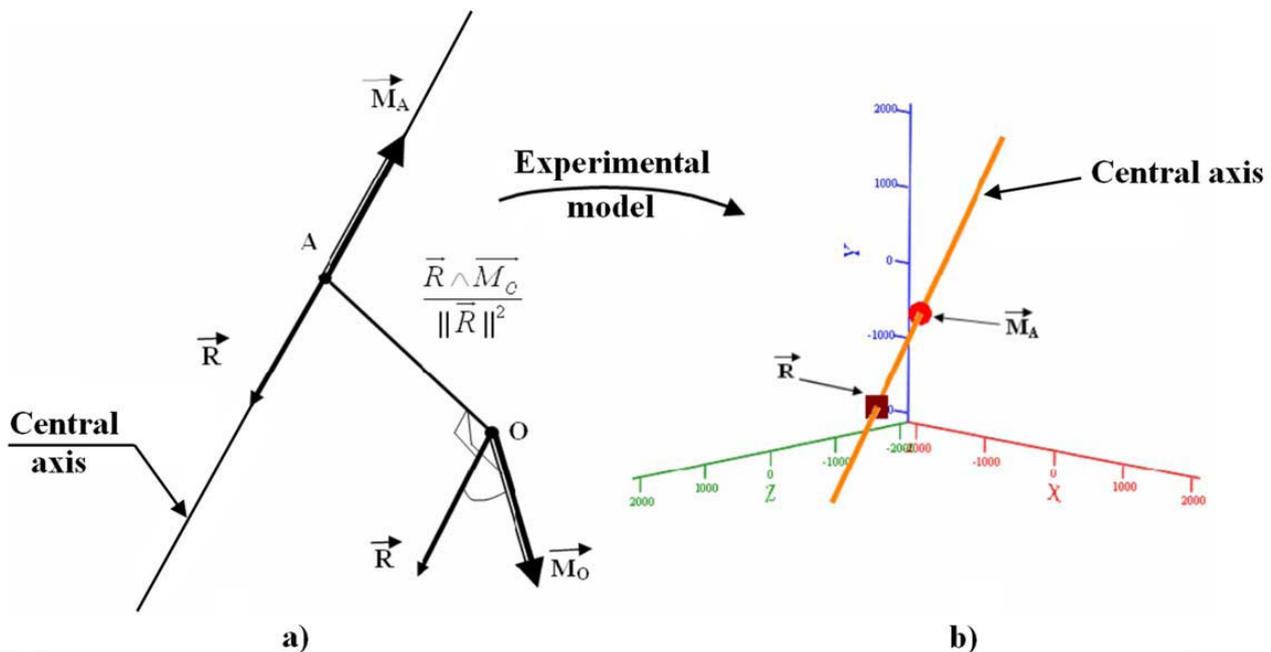

Figure [10]. Central axis representation (**a**) and of the colinearity between vector sum **R** and minimum moment $M_A$ at central axis (**b**).

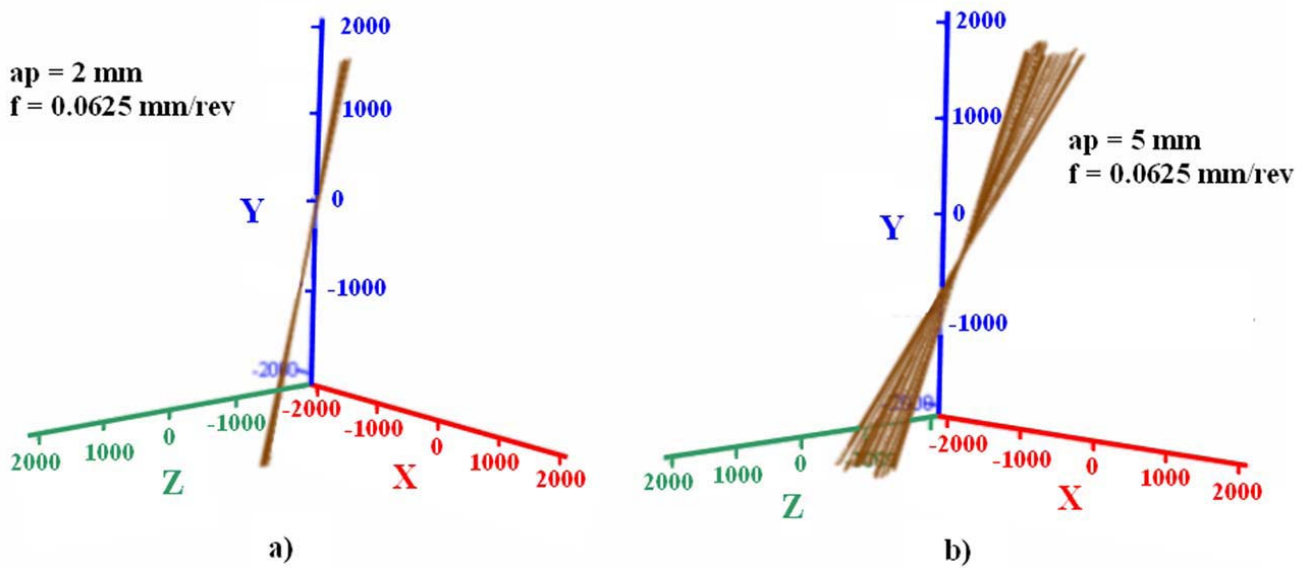

Figure [11]: Central axes representation obtained for 68 spin rounds of the workpiece speed and feed rate f = 0.0625 mm/rev; **a)** stable process ap = 2 mm; **b)** unstable process ap = 5 mm.

In the presence of vibrations (ap = 5 mm) for a 68 spin rounds of the workpiece (44 points of measurements per spin round), the dispersive character of the central axes beam, compared to the stable mode, can be observed, where this same beam is tightened more or less tilted compared to the normal axis on the plane (*x*, *y*). The self-excited vibrations, due to the variable moment generation, can explain this central axis dispersion.

## 5 Analysis of central axis moments related

While transporting the moment from the tool tip at the central axis, the minimum moment (pure torque) $M_A$ is obtained. From the moment values at the central axis, constant and variable parts are deduced. Just like for the efforts, the variable part is due to the self-excited vibrations, as revealed below.

Using this decomposition, the moments contribution on the areas of contact tool–workpiece–chip is expressed. The observations resulting from the analysis show that the tool vibrations generate rotations; cause variations of contact; and, thus, generate variable moments, confirming the efforts analysis detailed in Section 3. This representation allows us to express the moments along the three axes of the machine tool: swivel moment in the *y* direction and the two moments of rotation along *x* and *z* directions.

For the unstable case, characterized by ap = 5 mm and f = 0.1 mm/rev, the analysis of the cutting torque components average values at the central axis (subscripted by the letter a) leads to the following results: Max = - 4 dN.m ±24 dN.m; May = - 34 dN.m ±160 dN.m; Maz = 8 dN.m ±110 dN.m. Thus, one has the following relation between the modulus of the three torque components (with the errors of measurement close):

$$\|Max\| = \|Maz\|/2 = \|May\|/8. \qquad (10)$$

The confrontation of Eq (6) and (10) shows the importance of the moment transport from central axis to the tool tip point. It is as also interesting to note as the module of the average couple of cut at the central axis is worth $\|My\|$ = 88 dN.m, 3 times less than at the tool point.

In addition, one notes that the torque modulus is increasing with feed rate as for the resultant cutting force, at the central axis. The moment component along the radial axis (x) remains the weakest but also the least disturbed. Contrary to the moment components at the tool tip point or the resultant, the single order relations (3) and (8) are replaced by:

$$|Max| \leq |May| \leq |Maz| \quad \text{if } f \leq 0.0625, \qquad (11)$$

$$|Max| \leq |Maz| \leq |May| \quad \text{if } f > 0.0625. \qquad (12)$$

Depending on the feed rate, the role of cutting and feed rate axes is thus inversed. Moreover, the torque along the feed rate axis plays the role of "pivot", and keeps it of the transportation at the tool tip point for f > 0.0625. When f ≤ 0.0625, the role of "pivot" go back to the cutting axis, and the influence of cutting and the feed rate axes is inversed. This remark completes the previous one above, and highlights that the study of cutting moment at the central axis needs to examined in more details.

## 6 Conclusion

Experimental procedures developed allowed to determine elements required for a rigorous analysis of the influence of tool geometry, its displacement, and the evolution of contacts tool/workpiece and tool/chip on the machined surface. These experimental results allowed especially to establish a vectorial decomposition of actions by analysing the resultant of applied actions torsor during a turning process. Thus, it is demonstrated that the cutting effort varies around a constant face value, describing an "8" in a leaning plan compared with the machine spindle. This cutting force, whose application point describes an ellipse, is perfectly well correlated with displacements of the tool tip point, in a good agreement with [27].

Furthermore, a really innovative study of the moments at the tool tip point, and at the central axis of the actions torsor, is presented. For the resultant as for the moment, at the tool tip point or the central axis, the modulus of these elements is increasing with the feed rate. An order relation exists between the average values of modulus of action components and the average values of moments at tool tip point. One notes especially that the roles of x and y axes are inversed, and z axis plays the role of "pivot". But not unique order relation exists at the central axis. Only the moment modulus along the radial axis remains the weakest when the feed rate increases. This property is common with the component modulus of the same row of the resultant. For the module of the other components of the moments at the central axis, their relative position depends on the feed rate value. This highlights that the study of cutting moments at the central axis needs to be examined in more details.